\documentclass[namedreferences]{kluwer}    

\newcommand{\etal}{{\em et {al.}}}

\newcommand{\apj}{{\em Astrophys.\ J.}}
\newcommand{\apjl}{{\em Astrophys.\ J.\ (Letters)}}

\newcommand{\aap}{{\em Astron.\ Astrophys.}}

\newcommand{\mnras}{{\em Mon.\ Not.\ R.\ Astron.\ Soc.}}

\newcommand{\rmp}{{\em Rev.\ Mod.\ Phys.}}

\begin{document}                                                                                   
\begin{article}
\begin{opening}         
\title{Turbulence in the Interstellar Medium} 

\subtitle{Energetics and Driving Mechanisms}

\author{Mordecai-Mark \surname{Mac Low}}  
\runningauthor{Mordecai-Mark Mac Low}
\runningtitle{Turbulence in the Interstellar Medium}
\institute{Department of Astrophysics, American Museum of Natural
History\\ New York, NY 10024-5192, USA \email{mordecai@amnh.org}}

\begin{abstract}
Interstellar turbulence is expected to dissipate quickly in the
absence of continuous energy input.  I examine the energy available
for driving the turbulence from six likely mechanisms:
magnetorotational instability, gravitational instability, protostellar
outflows, H~{\sc ii} region expansion, stellar winds, and supernovae.
I conclude that supernovae contribute far more energy than the other
mechanisms, and so form the most likely drving mechanism.
\end{abstract}

\end{opening}           

\section{Introduction}

Turbulent flows appear more and more central to our understanding of
the interstellar medium.  The distribution of pressures and densities
are probably determined as much by turbulent ram pressures as by
thermal phase transitions (V\'azquez-Semadeni, Gazol, \& Scalo 2000;
Mac Low \etal\ 2002).  Furthermore, compression by large-scale
turbulent flows may form molecular clouds (Ballesteros-Paredes,
Hartmann, \& V\'azquez-Semadeni 1999), and drive turbulence within
those clouds (Ossenkopf \& Mac Low 2002) to support them against
gravitational collapse (Klessen, Heitsch, \& Mac Low 2000).

Both support against gravity and maintenance of observed motions
appear to depend on continued driving of the turbulence, which has
kinetic energy density $e = (1/2) \rho v_{\rm rms}^2$.  Mac Low
(1999, 2002) estimates that the dissipation rate for isothermal, supersonic
turbulence is 
\begin{eqnarray} \label{eqn:dissip}
\dot{e} & \simeq & -(1/2)\rho v_{\rm rms}^3/L_d  \nonumber \\
        & = & -(3 \times 10^{-27} \mbox{ erg cm}^{-3} \mbox{ s}^{-1}) n
\left(\frac{v_{\rm rms}}{10 \mbox{ km s}^{-1}}\right)^3 
\left(\frac{L_d}{100 \mbox{ pc}}\right)^{-1},
\end{eqnarray}
where $n$ is the number density in units of cm$^{-3}$, $L_d$ is the
driving scale, which we have somewhat arbitrarily taken to be 100~pc
(though it could well be smaller), and we have assumed a mean mass per
particle $\mu= 2.11\times 10^{-24}$~g.  The dissipation time for
turbulent kinetic energy
\begin{equation} \label{eqn:disstime}
\tau_d = e / \dot{e} \simeq L/v_{\rm rms} = (9.8 \mbox{ Myr})
\left(\frac{L_d}{100 \mbox{ pc}}\right)
\left(\frac{v_{\rm rms}}{10 \mbox{ km s}^{-1}}\right)^{-1},
\end{equation}
which is just the crossing time for the turbulent flow across the
driving scale (Elmegreen 2000b).  What then is the energy source for
this driving? We here review the energy input rates for a number of
possible mechanisms.

\section{Magnetorotational Instabilities}

One energy source for interstellar turbulence that has long been
considered is shear from galactic rotation (Fleck 1981).  However, the
question of how to couple from the large scales of galactic rotation
to smaller scales remained open.  Work by Sellwood \& Balbus (1999)
has shown that the magnetorotational instability (Balbus \& Hawley
1991, 1998) could couple the large-scale motions to small scales
efficiently.  The instability generates Maxwell stresses (a positive
correlation between radial $B_R$ and azimuthal $B_{\Phi}$ magnetic
field components) that transfer energy from shear into turbulent
motions at a rate $\dot{e} = T_{R\Phi} \Omega$ (Sellwood \& Balbus
1999).  Numerical models suggest that the Maxwell stress tensor
$T_{R\Phi} \simeq 0.6 B^2/(8\pi)$ (Hawley, Gammie \& Balbus 1995).
For the Milky Way, the value of the rotation rate recommended by the
IAU is $\Omega = (220 \mbox{ Myr})^{-1} = 1.4 \times 10^{-16} \mbox{ rad
s}^{-1}$, though this may be as much as 15\% below the true value
(Olling \& Merrifield 1998, 2000).  The magnetorotational instability
may thus contribute energy at a rate
\begin{equation}
\dot{e} = (3 \times 10^{-29} \mbox{ erg cm$^{-3}$ s}^{-1})
\left(\frac{B}{3 \mu\mbox{G}}\right)^2 \left(\frac{\Omega}{(220 \mbox{
Myr})^{-1}}\right). 
\end{equation}
For parameters appropriate to the H{\sc i} disk of a sample small
galaxy, NGC~1058, Sellwood \& Balbus (1999) find that the magnetic
field required to produce the observed velocity dispersion of
6~km~s$^{-1}$ is roughly 3 $\mu$G, a reasonable value for such a
galaxy.  This instability may provide a base value for the velocity
dispersion below which no galaxy will fall.  If that is sufficient to
prevent collapse, little or no star formation will occur, producing
something like a low surface brightness galaxy with large amounts of
H~{\sc i} and few stars.

\section{Gravitational Instabilities}

Motions coming from gravitational collapse have often been suggested
as a local driving mechanism in molecular clouds, but fail due to the
quick decay of the turbulence (Klessen, Burkert, \& Bate 1998).  If
the turbulence decays in less than a free-fall time, as suggested by
equation~\ref{eqn:dissip}, then it cannot delay collapse for
substantially longer than a free-fall time.

On the galactic scale, spiral structure can drive turbulence in gas
disks.  Roberts (1969) first demonstrated that shocks would form in
gas flowing through spiral arms formed by gravitational instabilities
in the stellar disk (Lin \& Shu 1964, Lin, Yuan, \& Shu 1969).  These
shocks were studied in thin disks by Tubbs (1980) and Soukup \& Yuan
(1981), who found few vertical motions.  More recently, it has been
realized that in a more realistic thick disk, the spiral shock will
take on some properties of a hydraulic bore, with gas passing through
a sudden vertical jump at the position of the shock (Martos \& Cox
1998, G\'omez \& Cox 2002).  Behind the shock, downward flows of as
much as 20~km~s$^{-1}$ appear (G\'omez \& Cox 2002).  Some portion of
this flow will contribute to interstellar turbulence.  However, the
observed presence of interstellar turbulence in irregular galaxies
without spiral arms, as well as in the outer regions of spiral
galaxies beyond the regions where the arms extend suggest that this
cannot be the only mechanism driving turbulence.  A more quantitative
estimate of the energy density contributed by spiral arm driving
has not yet been done.

The interaction between rotational shear and gravitation can, at least
briefly, drive turbulence in a galactic disk, even in the absence of
spiral arms. This process has been numerically modeled at high
resolution (sub-parsec zones) in two dimensions in a series of papers
by Wada \& Norman (1999, 2001), Wada, Spaans, \& Kim (2000), and Wada,
Meurer \& Norman (2002).  However, these models all share two
limitations: they do not include the dominant stellar component, and
gravitational collapse cannot occur beneath the grid scale. The
computed filaments of dense gas are thus artificially supported, and
would actually continue to collapse to form stars, rather than driving
turbulence in dense disks (see S\'anchez-Salcedo [2001] for a detailed
critique).  In very low density disks, where even the dense filaments
remained Toomre stable, this mechanism might operate, however.

Wada \etal\ (2002) estimated the energy input from this mechanism
following the lead of Sellwood \& Balbus (1999), but substituting
Newton stresses (Lynden-Bell \& Kalnajs 1972) for Maxwell stresses.
The Newton stresses will only add energy if a positive correlation
between radial and azimuthal gravitational forces exists, however,
which is not demonstrated by Wada \etal\ (2002).  Nevertheless, they
estimate the order of magnitude of the energy input from Newton
stresses as
\begin{eqnarray}
\dot{e} & \simeq & G (\Sigma_g/H) \lambda^2 \Omega \nonumber \\
        & \simeq & (4 \times 10^{-29} \mbox{ erg cm$^{-3}$ s}^{-1})
\left(\frac{\Sigma_g}{10 \mbox{ M$_{\odot}$ pc}^{-2}} \right)^2 \times
        \nonumber \\         & \times & 
\left(\frac{H}{100 \mbox{ pc}} \right)^{-2}
\left(\frac{\lambda}{100 \mbox{ pc}} \right)^{2}
\left(\frac{\Omega}{(220 \mbox{ Myr})^{-1}}\right), 
\end{eqnarray}
where $G$ is the gravitational constant, $\Sigma_g$ the density of
gas, $H$, the scale height of the gas, $\lambda$ a length scale of
turbulence, and $\Omega$ the angular velocity of the disk.  Values
chosen are appropriate for the Milky Way.  This is two orders of
magnitude below the value required to maintain interstellar turbulence
(eq.~[\ref{eqn:dissip}]).

\section{Protostellar outflows}

Protostellar jets and outflows are a popular suspect for the energy
source of the observed turbulence.  We can estimate their average
energy input rate, following McKee (1989), by assuming that some
fraction $f_w$ of the mass accreted onto a star during its formation
is expelled in a wind travelling at roughly the escape velocity.  Shu
\etal\ (1988) argue that $f_w \simeq 0.4$, and that most of the mass
is ejected from close to the stellar surface, where the escape
velocity 
\begin{equation}
v_{\rm esc} = (2GM/R)^{1/2} = (200 \mbox{ km s}^{-1}) (M/\mbox{
M}_{\odot})^{1/2} (R/10\mbox{ R}_{\odot})^{-1/2},
\end{equation}
where the scaling is appropriate for a protostar with mass $M = 1
\mbox{ M}_{\odot}$ and radius $R = 10 \mbox{ R}_{\odot}$.
Observations of neutral atomic winds from protostars suggest outflow
velocities of roughly this value (Lizano \etal\ 1988, Giovanardi et
al.\ 2000).

The total energy input from protostellar winds will substantially
exceed the amount that can be transferred to the turbulence due to
radiative cooling at the wind termination shock.  We represent the
fraction of energy lost there by $\eta_w$.  A reasonable upper limit
to the energy loss is offered by assuming fully effective radiation
and momentum conservation, so that 
\begin{equation}
\eta_w < \frac{v_{\rm rms}}{v_w} = 0.05 \left(\frac{v_{\rm rms}}{10 \mbox{ km
s}^{-1}}\right) \left(\frac{200 \mbox{ km s}^{-1}}{v_w}\right),
\end{equation}
where $v_{\rm rms}$ is the rms velocity of the turbulence, and we have
assumed that the flow is coupled to the turbulence at typical
velocities for the diffuse ISM.  If we assumed that most of the energy
went into driving dense gas, the efficiency would be lower, as typical
rms velocities for CO outflows are 1--2~km~s$^{-1}$. The energy
injection rate 
\begin{eqnarray}
\dot{e}& = & \frac12 f_w \eta_w \frac{\dot{\Sigma}_*}{H} v_w^2
       \nonumber \\
       & \simeq &  (2 \times 10^{-28} \mbox{ erg cm}^{-3} \mbox{ s}^{-1})
       \left(\frac{H}{200 \mbox{ pc}}\right)^{-1} 
       \left(\frac{f_w}{0.4}\right)       
       \left(\frac{v_{\rm rms}}{10 \mbox{ km s}^{-1}}\right)  \times
 \nonumber \\
& \times  &    \left(\frac{v_w}{200 \mbox{ km s}^{-1}}\right) 
       \left(\frac{v_{\rm rms}}{10 \mbox{ km s}^{-1}}\right) 
       \left(\frac{\dot{\Sigma}_*}{4.5 \times 10^{-9} \mbox{ M$_{\odot}$
       pc$^{-2}$ yr$^{-1}$}}\right),
\end{eqnarray}
where $\dot{\Sigma}_*$ is the surface density of star formation, and
$H$ is the scale height of the star-forming disk.  The scaling value
used for $\dot{\Sigma}_*$ is the solar neighborhood value (McKee 1989).

Although protostellar jets and winds are indeed quite energetic, they
deposit most of their energy into low density gas (Henning 1989), as
is shown by the observation of multi-parsec long jets extending
completely out of molecular clouds (Bally \& Devine 1994).
Furthermore, observed motions of molecular gas show increasing power
on scales all the way up to and perhaps beyond the largest scale of
molecular cloud complexes (Ossenkopf \& Mac Low 2002).  It is hard to
see how such large scales could be driven by protostars embedded in
the clouds.

\section{Massive Stars}
In active star-forming galaxies, however, massive stars appear likely
to dominate the driving.  They do so through ionizing radiation and
stellar winds from O~stars, and clustered and field supernova
explosions, predominantly from B~stars no longer associated with their
parent gas.  The supernovae appear likely to dominate, as we now show.

\subsection{Stellar Winds}
First, we consider stellar winds.  The total energy input from a
line-driven stellar wind over the main-sequence lifetime of an early
O~star can equal the energy from its supernova explosion, and the
Wolf-Rayet wind can be even more powerful.  However, the mass-loss
rate from stellar winds drops as roughly the sixth power of the star's
luminosity if we take into account that stellar luminosity varies as
the fourth power of stellar mass (Vink, de Koter \& Lamers 2000), and
the powerful Wolf-Rayet winds (Nugis \& Lamers 2000) last only $10^5$
years or so, so only the very most massive stars contribute
substantial energy from stellar winds.  The energy from supernova
explosions, on the other hand, remains nearly constant down to the
least massive star that can explode.  As there are far more lower-mass
stars than massive stars, with a Salpeter (1955) IMF giving a
power-law in mass of $\alpha = -2.35$, supernova explosions will
inevitably dominate over stellar winds after the first few million
years of the lifetime of an OB association, until the lifetime of the
least massive star to explode of around 40--50 Myr.

\subsection{H{\sc ii} Region Expansion}
Next, we consider ionizing radiation from OB stars.  The total
amount of energy contained in ionizing radiation is vast.  Abbott
(1982) estimates the total luminosity of ionizing radiation in the
disk to be 
\begin{equation}
\dot{e} = 1.5 \times 10^{-24} \mbox{ erg s$^{-1}$ cm}^{-3}.
\end{equation}
However, only a very small fraction of this total energy goes to driving
interstellar motions, as we now show.

Ionizing radiation primarily contributes to interstellar turbulence by
ionizing H{\sc ii} regions, heating them to 7000--10,000~K, and raising
their pressures above that of surrounding neutral gas, so that they
expand supersonically.  Matzner (2002) computes the momentum input
from the expansion of an individual H{\sc ii} region into a
surrounding molecular cloud, as a function of the cloud mass and the ionizing
luminosity of the central OB association.  By integrating over the
H{\sc ii} region luminosity function derived by McKee \& Williams
(1997), he finds that the average momentum input from a Galactic
region is 
\begin{equation}
\langle \delta p \rangle \simeq (260 \mbox{ km s}^{-1})
\left(\frac{N_H}{1.5 \times 10^{22} \mbox{ cm}^{-2}}\right)^{-3/14}
\left(\frac{M_{cl}}{10^6 \mbox{ M}_{\odot}}\right)^{1/14} 
\langle M_* \rangle.
\end{equation}
The column density $N_H$ is scaled to the mean value for Galactic
molecular clouds (Solomon \etal\ 1987), which varies little as cloud
mass $M_{cl}$ varies.  The mean stellar mass per cluster in the Galaxy
$\langle M_* \rangle = 440 \mbox{ M}_{\odot}$ (Matzner 2002).

The number of OB associations contributing substantial amounts of
energy can be drawn from the McKee \& Williams (1997) cluster
luminosity function
\begin{equation}
{\cal N} (> S_{49}) = 6.1 \left(\frac{108}{S_{49}} - 1 \right),
\end{equation}
where ${\cal N}$ is the number of associations with ionizing photon
luminosity exceeding $S_{49} = S/(10^{49}$~s$^{-1})$.  The
luminosity function is rather flat below $S_{49} = 2.4$, the
luminosity of a single star of 120~M$_{\odot}$, which was the highest
mass star considered, so taking its value at $S_{49} = 1$ is about
right, giving ${\cal N}(> 1) = 650$ clusters.

To derive an energy input rate per unit volume $\dot{e}$ from the mean
momentum input per cluster $\langle \delta p \rangle$, we need to
estimate the average velocity of momentum input $v_{i}$, the time over
which it occurs $t_{i}$, and the volume $V$ under consideration.
Typically expansion will not occur supersonically with respect to the
interior, so $v_{i} < c_{s,i}$, where $c_{s,i} \simeq 10$~km~s$^{-1}$
is the sound speed of the ionized gas.  McKee \& Williams (1997) argue
that clusters typically last for about five generations of massive
star formation, where each generation lasts $\langle t_* \rangle =
3.7$~Myr. The scale height for massive clusters is $H_c \sim 100$~pc
(e.g.\ Bronfman \etal\ 2000), and the radius of the star-forming disk
is roughly $R_{sf} \sim 15$~kpc, so the relevant volume $V = 2 \pi
R_{sf}^2 H_c$.  The energy input rate from H{\sc ii} regions is then
\begin{eqnarray}
\dot{e}& =& \frac{\langle \delta p \rangle {\cal N}(>1) v_{i}}{V
t_{i}} \nonumber \\
       & = & (3 \times 10^{-30} \mbox{ erg s$^{-1}$ cm}^{-3})
\left(\frac{N_H}{1.5\times 10^{22} \mbox{ cm}^{-2}}\right)^{-3/14}
\times \nonumber \\   & \times &
\left(\frac{M_{cl}}{10^6 \mbox{ M}_{\odot}}\right)^{1/14}
\left(\frac{\langle M_* \rangle}{440 \mbox{ M}_{\odot}}\right)
\left(\frac{{\cal N}(>1)}{650}\right) 
\times \nonumber \\ & \times &  
\left(\frac{v_{i}}{10 \mbox{ km s}^{-1}}\right)
\left(\frac{H_c}{100 \mbox{ pc}}\right)^{-1}
\left(\frac{R_{sf}}{15 \mbox{ kpc}}\right)^{-2}
\left(\frac{t_{i}}{18.5 \mbox{ Myr}}\right)^{-1}, 
\end{eqnarray}
where all the scalings are appropriate for the Milky Way as discussed
above.  Nearly all of the energy in ionizing radiation goes towards
maintaining the ionization of the diffuse ionized medium, and hardly
any towards driving turbulence.  Flows of ionized gas may be important
very close to young clusters, but do not appear to contribute
significantly on a global scale.

\subsection{Supernovae}
The largest contribution from massive stars to interstellar turbulence
comes from supernova explosions.  To estimate their energy input rate,
we begin by finding the supernova rate in the Galaxy $\sigma_{SN}$.
Cappellaro et al. (1999) estimate the total supernova rate in
supernova units to be $0.72 \pm 0.21$ SNu for galaxies of type S0a-b
and $1.21 \pm 0.37$~SNu for galaxies of type Sbc-d, where 1 SNu = 1 SN
(100~yr)$^{-1} (10^{10} L_B/\mbox{L}_{\odot})^{-1}$, and $L_B$ is the
blue luminosity of the galaxy.  Taking the Milky Way as lying between
Sb and Sbc, we estimate $\sigma_{SN} = 1$~SNu.  Using a Galactic
luminosity of $L_B = 2 \times 10^{10} \mbox{ L}_{\odot}$, we find a
supernova rate of (50~yr)$^{-1}$, which agrees well with the estimate
in equation~(A4) of McKee (1989).  If we use the same scale height
$H_c$ and star-forming radius $R_{sf}$ as above, we can compute the
energy input rate from supernova explosions with energy $E_{SN} =
10^{51}$~erg to be
\begin{eqnarray}
\dot{e} & = &\frac{\sigma_{SN} \eta_{SN} E_{SN}}{\pi R_{sf}^2 H_c}
       \nonumber \\
       &  = & (3 \times 10^{-26} \mbox{ erg s$^{-1}$ cm}^{-3})
\left(\frac{\eta_{SN}}{0.1} \right)
\left(\frac{\sigma_{SN}}{1 \mbox{ SNu}} \right) 
\times \nonumber \\ & \times &
\left(\frac{H_c}{100 \mbox{ pc}} \right)^{-1} 
\left(\frac{R_{sf}}{15 \mbox{ kpc}} \right)^{-2}
\left(\frac{E_{SN}}{10^{51} \mbox{ erg}} \right).
\end{eqnarray}
The efficiency of energy transfer from supernova blast waves to the
interstellar gas $\eta_{SN}$ depends on the strength of radiative
cooling in the initial shock, which will be much stronger in the
absence of a surrounding superbubble (e.g.\ Heiles 1990).  Substantial
amounts of energy can escape in the vertical direction in superbubbles
as well, however.  (Norman \& Ferrara [1996] make an analytic estimate
of the effectiveness of driving by SN remnants and superbubbles.) The
scaling factor $\eta_{SN} \simeq 0.1$ used here was derived by
Thornton \etal\ (1998) from one-dimensional numerical simulations of
SNe expanding in a uniform ISM, or can alternatively be drawn from
momentum conservation arguments comparing a typical expansion velocity
of 100~km~s$^{-1}$ to typical interstellar turbulence velocity of
10~km~s$^{-1}$.  Detailed multi-dimensional modeling of the
interactions of multiple SN remnants (e.g.\ Avillez 2000) will be
required to better determine it.

Supernova driving appears to be powerful enough to maintain the
turbulence even with the dissipation rates estimated in
equation~(\ref{eqn:dissip}).  It provides a large-scale
self-regulation mechanism for star formation in disks with sufficient
gas density to collapse despite the velocity dispersion produced by
the magnetorotational instability.  As star formation increases in
such galaxies, the number of OB stars increases, ultimately increasing
the supernova rate and thus the velocity dispersion, which restrains
further star formation.

Supernova driving not only determines the velocity dispersion, but may
actually form molecular clouds by sweeping gas up in a turbulent
flow. Clouds that are turbulently supported will experience
inefficient, low-rate star formation, while clouds that are too
massive to be supported will collapse (e.g. Kim \& Ostriker 2001),
undergoing efficient star formation to form OB associations or even
starburst knots.





\section{Conclusions}

Interstellar gas is only quiescent on the very smallest scales
($<0.1$~pc) in dense cores.  The transsonic and supersonic flows
observed elsewhere must be driven, as turbulent motions otherwise
decay away on timescales of order the crossing time of their outer
scale.  The largest energy source in star-forming regions of galaxies
is supernova explosions.  Even turbulence within dense molecular
clouds may derive primarily from field supernovae (not the few
supernovae that may be associated with the cloud itself before its
disappearance, but the background explosions from B stars formed
within the previous 50~Myr or so.)  Outside of star-forming parts of
galactic disks, turbulence may be driven by magnetorotational
instabilities, or even interaction between gravitation and rotation.

\acknowledgements I thank the organizers of the conference for their
invitation and partial support of my attendance.  This talk was
adapted from a section of Mac Low \& Klessen (2003).  This research
was partially supported by US NSF CAREER grant AST99-85392.

\theendnotes

\end{article}
\end{document}